\documentclass[CMYK,NoSecthm,linkcolor=black,citecolor=black,colorlinks=false,bookmarksnumbered]{jcomsec_author}

\usepackage{savesym}
\savesymbol{AND}
\usepackage{xcolor}
\usepackage{float}
\newcommand{\rev}[1]{{\color{black}#1}}
\usepackage{multicol}
\usepackage{enumerate}
\usepackage{amssymb}
\usepackage{multirow}
\usepackage{amsmath}
\usepackage{graphicx}
\usepackage{subcaption}
\usepackage{setspace}
\usepackage{mathtools}
\usepackage{colortbl}
\usepackage{color}
\usepackage{ctable}
\usepackage{wrapfig}
\usepackage[numbers,sort&compress]{natbib}
\usepackage{hypernat}
\usepackage{array}
\usepackage{microtype}
\usepackage{doi}
\usepackage{algorithm}
\usepackage{algorithmic}
\usepackage{url}

\usepackage[all]{hypcap}

\DisableLigatures[f]{encoding = *, family = *}
\newcolumntype{X}[1]{>{\centering\arraybackslash}m{#1}}

\begin{document}

\makeatletter
\renewcommand{\theHsection}{\arabic{section}}
\renewcommand{\theHsubsection}{\arabic{subsection}}
\renewcommand{\theHsubsubsection}{\arabic{subsubsection}}
\renewcommand{\theHfigure}{\arabic{figure}}
\renewcommand{\theHtable}{\arabic{table}}
\renewcommand{\theHalgorithm}{\arabic{algorithm}}

\@ifundefined{theHpart}{}{%
  \renewcommand{\theHpart}{\arabic{part}}%
}
\makeatother

\begin{frontmatter}

\title{Enhancing Error Detection Performance through Parallel CRC Computation on Multi-Core Architectures}

\author{Mohammadjavad Khani$^{~a}$}
\author{Mahmood Ahmadi$^{~a,*}$}
\thanks{$^{*}$ Mahmood Ahmadi.\\
Email addresses: m.khani@razi.ac.ir (M. Khani), 
m.ahmadi@razi.ac.ir (M. Ahmadi)}

\address[a]{Computer Engineering and Information Technology Department, Razi University, Kermanshah, Iran}

\runtitle{}
\runauthor{M. Khani and M. Ahmadi}
\begin{abstract}
 Cyclic Redundancy Check (CRC) remains one of the most widely used error-detection mechanisms in communication, storage, and embedded systems. However, conventional software CRC implementations suffer from inherent sequential dependencies that limit efficient utilization of modern multi-core processors. This paper presents a generalized software-based parallel CRC framework for multi-core architectures using POSIX threads (pthreads). The proposed framework supports multiple CRC variants, including CRC-8, CRC-16, CRC-32, CRC-64, and CRC-128, within a unified implementation model. {\color{black} To preserve correctness during parallel execution, the framework employs a GF(2)-based CRC combination mechanism rather than na\"ive XOR aggregation. The combine stage is formulated using polynomial arithmetic and matrix-based shifting operations over GF(2), ensuring equivalence between parallel and serial CRC computation. The proposed method was evaluated using multiple workload sizes and thread configurations. Experimental analysis includes execution time, throughput, latency, scalability behavior, and energy estimation under varying thread counts. Results indicate that parallel execution significantly improves performance for large datasets, achieving approximately 3–4× speedup on the evaluated platform while preserving exact CRC correctness. Comparative discussion with representative CRC optimization approaches, including lookup-table methods, slicing-by-8, SIMD/vectorized CRC, and hardware-assisted CRC techniques, is also provided to position the proposed framework within the broader CRC optimization landscape. Although the presented framework emphasizes portability, multi-variant support, and correctness-preserving software parallelization, scalability remains influenced by synchronization overhead, memory bandwidth constraints, and hardware characteristics. Overall, the proposed approach provides a portable and generalized software framework for correctness-preserving parallel CRC acceleration on general-purpose multi-core systems.}

\end{abstract}


\end{frontmatter}

\addtolength{\parskip}{2mm}

\section{Introduction}
{\color{black} Reliable error detection remains a fundamental requirement in modern digital communication, storage, networking, and embedded computing systems. During transmission or storage, data may be corrupted by noise, hardware faults, electromagnetic interference, or synchronization errors. Consequently, efficient mechanisms for detecting accidental bit modifications are essential for maintaining data integrity.}

Among various error-detection techniques, Cyclic Redundancy Check (CRC) has become one of the most widely adopted methods because of its strong error-detection capability, low implementation complexity, and flexibility across different application domains [1,22]. CRC mechanisms are extensively used in communication protocols, storage devices, Ethernet networks, wireless systems, embedded controllers, and file integrity verification frameworks [21].

CRC computation is traditionally implemented through polynomial arithmetic over the finite field GF(2) [1,22]. Although CRC algorithms are computationally efficient, conventional software implementations exhibit an inherent sequential dependency. The CRC state generated for a given data element depends directly on the previously computed state, creating a recursive processing chain. As a result, straightforward exploitation of modern multi-core processors becomes challenging.

{\color{black} The increasing prevalence of multi-core and many-core computing platforms motivates renewed interest in software parallelization strategies for CRC computation. Contemporary processors provide substantial thread-level parallelism capabilities that can potentially accelerate CRC workloads for large datasets. However, achieving correct CRC parallelization is nontrivial. A na\"ive division of the input stream followed by direct XOR aggregation of partial CRC values does not preserve the mathematical properties of CRC concatenation because CRC results depend on message ordering and polynomial state advancement [4,7].}

{\color{black} Several optimization techniques have been proposed to improve CRC performance, including lookup-table approaches, slicing-by-8 methods, SIMD/vectorized implementations, carry-less multiplication techniques, and hardware-assisted CRC instructions [10,11,14]. While these approaches provide substantial acceleration in specific environments, they frequently focus on particular CRC variants, hardware features, or instruction-set support.}

{\color{black} In addition, many existing software studies emphasize CRC-32 exclusively, leaving broader multi-variant implementations comparatively underexplored [14,18]. Furthermore, practical considerations such as portability, correctness-preserving combine operations, scalability behavior, and energy-oriented evaluation remain important for software implementations intended for general-purpose processors.}

{\color{black} This paper presents a generalized software-based framework for parallel CRC computation using POSIX threads (pthreads). The proposed approach supports multiple CRC families, including CRC-8, CRC-16, CRC-32, CRC-64, and CRC-128, under a unified implementation model. To preserve mathematical correctness during parallel execution, the framework employs a GF(2)-based CRC combination mechanism derived from polynomial arithmetic and matrix-based shifting operations rather than na\"ive XOR aggregation [12,13].}

{\color{black} The proposed framework is evaluated using execution time, throughput, latency, scalability behavior, and energy-oriented measurements under different workload sizes and thread configurations. The study additionally discusses synchronization overhead, memory bandwidth effects, and practical scalability limitations associated with software CRC acceleration on multi-core systems.}

{\color{black} The main contributions of this work can be summarized as follows:}

\begin{enumerate}
\item {\color{black} A generalized pthread-based software framework supporting multiple CRC variants, including CRC-8, CRC-16, CRC-32, CRC-64, and CRC-128.}

\item {\color{black} A correctness-preserving GF(2)-based CRC combination mechanism for parallel chunk aggregation.}

\item {\color{black} Extended experimental evaluation incorporating throughput, latency, scalability analysis, and energy-oriented measurements.}

\item {\color{black} Discussion of practical implementation trade-offs involving synchronization overhead, memory behavior, and multi-core scalability considerations.}
\end{enumerate}

The remainder of this paper is organized as follows. Section 2 reviews and analyzes existing and related works. Section 3 introduces the mathematical background and implementation methodology. Section 4 presents the proposed parallel CRC framework and GF(2)-based combine mechanism. Section 5 reports experimental evaluation and discussion. Finally, Section 6 concludes the paper and outlines future research directions.

\begin{table*}[t]
\centering
\caption{\textcolor{black}{Representative CRC Optimization Approaches}}
\label{tab:relatedwork}

\small
\setlength{\tabcolsep}{4pt}
\renewcommand{\arraystretch}{1.15}

\begin{tabular*}{\textwidth}{@{\extracolsep{\fill}}l|c|c|c|c|c}
\hline
\textbf{Approach} &
\textbf{Software} &
\textbf{Parallel} &
\textbf{Multi-CRC Support} &
\textbf{Hardware Dependency} &
\textbf{Correct Combine} \\
\hline
Bitwise CRC & Yes & No & Partial & No & Yes \\
Lookup-Table CRC & Yes & No & Partial & No & Yes \\
Slicing-by-8 & Yes & Limited & Mainly CRC-32 & No & Yes \\
SIMD / Vectorized CRC & Yes & Yes & Limited & Processor-Specific & Yes \\
Hardware CRC Instructions & Partial & Yes & Limited & High & Yes \\
FPGA / Hardware CRC & No & Yes & Variable & High & Yes \\
\textbf{Proposed Framework} & \textbf{Yes} & \textbf{Yes} & \textbf{Yes} & \textbf{No} & \textbf{Yes} \\
\hline
\end{tabular*}
\end{table*}

\section{Related Works}

{\color{black} CRC optimization has been extensively studied across software, hardware, networking, storage, and embedded computing environments because of the widespread use of CRC in error detection applications [1,20,22].}

{\color{black} Early software CRC implementations primarily relied on bitwise polynomial division methods [1]. Although conceptually simple and mathematically straightforward, bitwise implementations often suffer from relatively high computational cost because each input bit must be processed sequentially through repeated shift and XOR operations.}

{\color{black} To improve software performance, lookup-table approaches were introduced [11]. Table-driven CRC methods precompute polynomial remainder values to reduce runtime computational overhead. Variants such as byte-wise lookup and slicing-by-4/slicing-by-8 techniques significantly accelerate CRC processing by exploiting precomputed transformation tables and wider memory accesses [11,14]. These approaches remain widely used in practical software libraries because of their favorable balance between simplicity and performance.}

{\color{black} Further acceleration has been achieved through instruction-level optimization methods. SIMD (Single Instruction Multiple Data) implementations leverage vector processing instructions such as SSE and AVX to process multiple data elements concurrently. In addition, carry-less multiplication techniques and dedicated hardware CRC instructions available on modern processors have demonstrated substantial performance improvements for selected CRC variants, particularly CRC-32 [10,14]. However, such approaches often depend on specific processor architectures, instruction-set extensions, or hardware capabilities.}

{\color{black} Several software libraries and operating system kernels incorporate optimized CRC combination mechanisms for concatenated data processing. For example, GF(2)-based combine operations have been employed in implementations such as crc32-combine to correctly merge CRC states derived from segmented message processing [12,13]. These approaches recognize that CRC values cannot be combined through na\"ive XOR aggregation because CRC correctness depends on polynomial state advancement and message ordering [11].}

{\color{black} Parallel CRC computation has also received attention in prior literature. Existing studies have explored multi-threaded CPU implementations, FPGA-based architectures, and specialized hardware pipelines to increase throughput for large-scale datasets and high-speed communication environments [2--9,15,16,18]. Hardware-oriented solutions frequently achieve very high acceleration; however, they may require specialized development environments or reduced portability compared with general-purpose software implementations.}

{\color{black} Despite significant prior progress, several practical gaps remain. Many published studies concentrate primarily on CRC-32 and do not examine a broader family of CRC variants within a unified framework [14,18]. Furthermore, some approaches prioritize raw throughput optimization while providing limited discussion of correctness-preserving combination mechanisms, scalability behavior, synchronization overhead, portability considerations, or energy-oriented evaluation.}

{\color{black} In contrast to hardware-specific acceleration methods, the present work focuses on a generalized software-oriented approach for multi-core processors using POSIX threads (pthreads). The proposed framework supports multiple CRC variants—including CRC-8, CRC-16, CRC-32, CRC-64, and CRC-128—within a unified implementation structure. A GF(2)-based combine formulation is employed to preserve equivalence between serial and parallel execution [12,13], and the experimental evaluation incorporates execution time, throughput, latency, scalability behavior, and energy-oriented analysis.}

{\color{black} Table 1 summarizes representative categories of existing CRC optimization approaches and positions the proposed framework relative to prior work. As shown in Table 1, the proposed framework is not intended to compete directly with highly specialized hardware-assisted CRC engines. Instead, it emphasizes portability, generalized multi-variant support, correctness-preserving parallel combination, and software-oriented evaluation on general-purpose multi-core platforms.}

\begin{figure}[t]
\centering
\includegraphics[width=\linewidth]{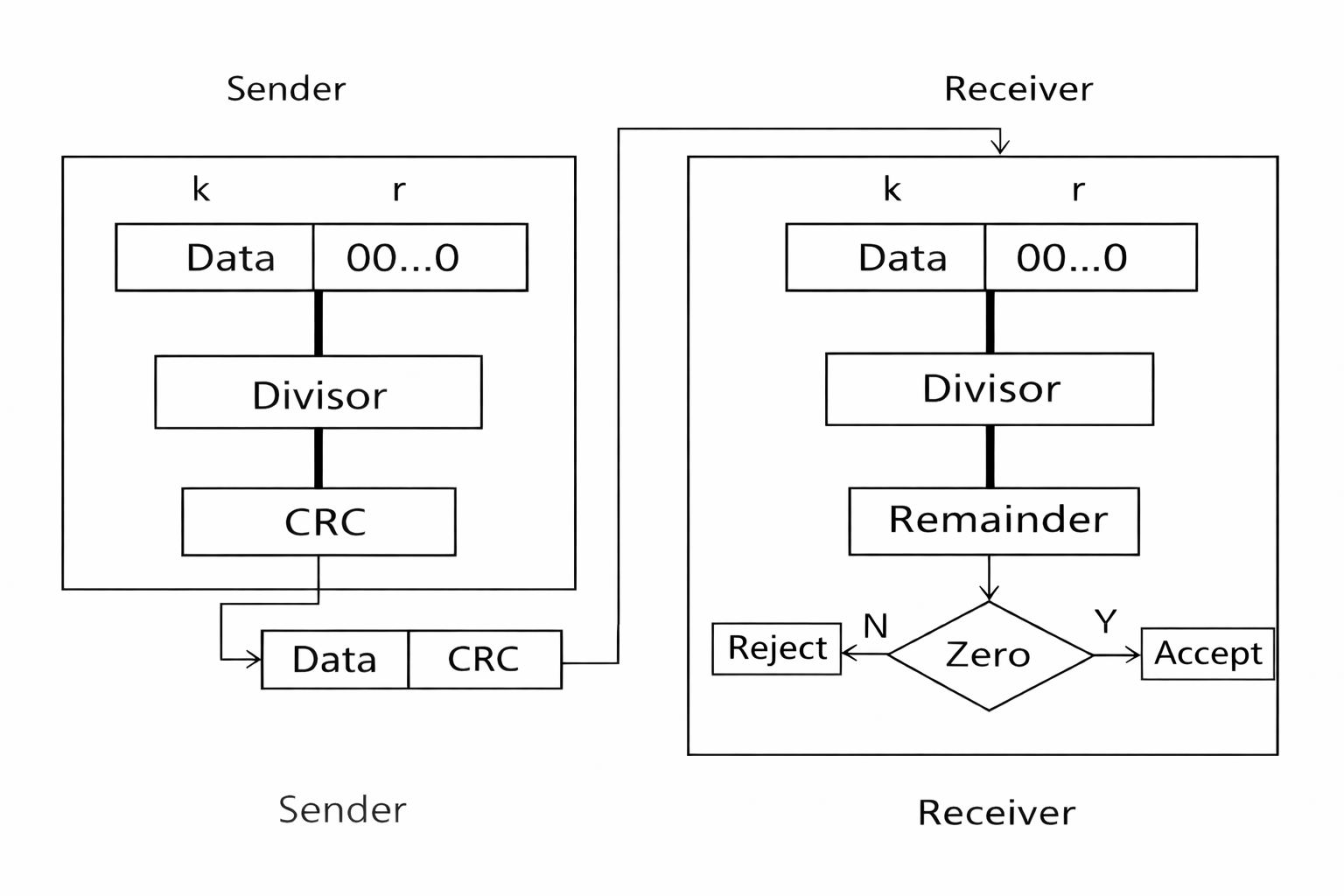}
\caption{Block Diagram of cyclic redundancy check [21].}
\label{fig:crc-block}
\end{figure}

\begin{figure}[t]
\centering
\includegraphics[width=\linewidth]{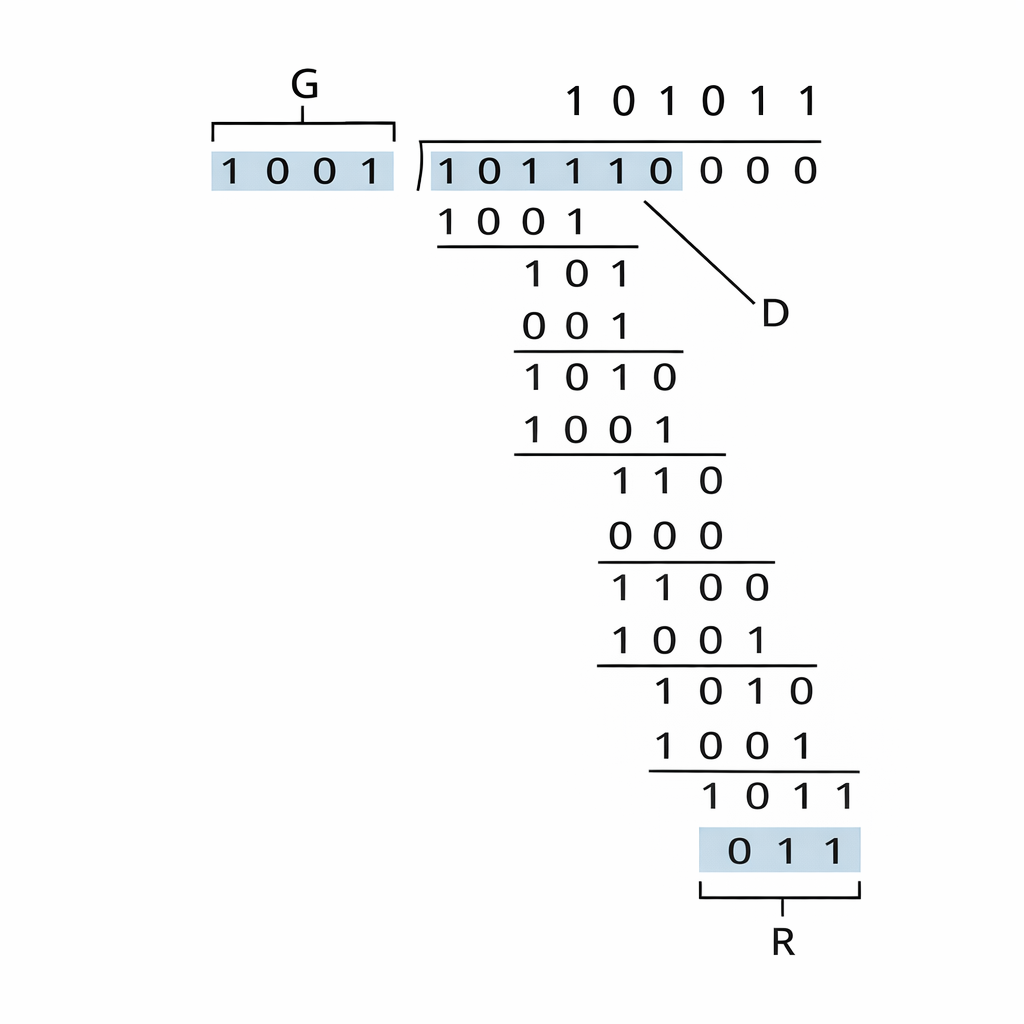}
\caption{A sample CRC calculation [20].}
\label{fig:crc-sample}
\end{figure}

\section{Fundamentals of Cyclic Redundancy Check (CRC)}

Cyclic Redundancy Check (CRC) is a widely used error detection mechanism designed to identify accidental changes in digital data during transmission or storage. CRC is based on polynomial arithmetic over the Galois Field GF(2), where both the data stream and the generator polynomial are represented as binary polynomials. The CRC value is computed by appending a fixed number of zero bits to the original data and performing a polynomial division by a predefined generator polynomial as shown in Fig. 1. The remainder of this division constitutes the CRC checksum, which is appended to the transmitted data. At the receiver side, the same division is performed to verify data integrity. As shown in Fig. 2, a sample CRC calculation is illustrated. The strength of CRC lies in its ability to detect common error patterns such as single-bit errors, burst errors, and multiple-bit errors, depending on the choice of the generator polynomial. Due to its strong detection capability and low redundancy overhead, CRC has been adopted as a standard error detection technique in numerous communication protocols and storage systems [1,21,22].

From an implementation perspective, CRC can be realized in both hardware and software, each offering distinct advantages. Hardware implementations often rely on linear feedback shift registers (LFSRs), enabling high-speed and low-latency CRC computation, which is particularly suitable for network interfaces and embedded controllers [2,3,8]. Software-based CRC implementations, on the other hand, provide greater flexibility and portability across different platforms. In software, CRC is commonly implemented using bitwise operations, lookup tables, or a combination of both. Bitwise implementations are simple and memory-efficient but tend to be computationally expensive, whereas table-driven implementations trade memory usage for higher throughput by precomputing partial CRC values [11,12,13]. The choice of implementation strategy depends on system constraints such as available memory, required throughput, and power consumption.

Performance considerations play a crucial role in CRC computation, especially in high-speed and data-intensive applications. The execution time of CRC algorithms is influenced by factors such as input data size, processor architecture, memory access patterns, and instruction-level parallelism. On modern processors, cache efficiency and branch prediction can significantly affect CRC performance. While table-based approaches reduce computational complexity, they may suffer from cache misses when large tables are used. Conversely, bitwise implementations are compute-bound and may limit throughput on large data streams. As data rates and data volumes continue to increase, optimizing CRC performance has become a critical challenge in software systems, particularly those operating under real-time or energy-constrained conditions [10,11,14,20].

With the widespread adoption of multi-core processors, parallel computing has emerged as an effective solution to overcome the performance limitations of traditional serial CRC implementations. Parallel CRC techniques divide input data into independent segments that can be processed concurrently, thereby exploiting thread-level parallelism to improve computational efficiency [2,4,7,15,16,18]. However, extending parallelization beyond a single CRC type introduces additional challenges, as different CRC variants (e.g., CRC-8, CRC-16, CRC-32, CRC-64, and CRC-128) exhibit distinct polynomial structures and computational complexities. Efficient parallel designs must therefore address workload balancing, synchronization overhead, and the correct combination of partial CRC results while preserving algorithmic correctness across multiple CRC formats [4,7,17,18]. A comprehensive understanding of the computational characteristics and structural differences among various CRC algorithms is thus essential for motivating, designing, and systematically evaluating the proposed parallel CRC framework presented in this paper.

\section{Proposed Parallel CRC Framework}

This section presents the proposed generalized software framework for parallel CRC computation on multi-core processors. The framework combines thread-level parallelism with a mathematically correct GF(2)-based combination mechanism to preserve equivalence with serial CRC execution.

\subsection{Threading Model and Static Data Partitioning}
{\color{black}

The proposed implementation employs POSIX threads (pthreads) because of their portability, deterministic synchronization behavior, and widespread availability across general-purpose operating systems. Given an input dataset $D$ of length $L$ bytes and a thread count $T$, the input message is partitioned into contiguous chunks:
\begin{equation}
D=D_0||D_1||D_2||\cdots||D_{T-1}
\end{equation}
where:
\begin{equation}
|D_i|\approx \frac{L}{T}
\end{equation}

Each thread independently computes the CRC value of its assigned partition.

Static partitioning is adopted to reduce runtime scheduling overhead, improve cache locality, and maintain predictable workload distribution. When the input size is not divisible by the number of threads, the remaining bytes are assigned to the final thread.

The chunk boundaries are computed as:
\begin{equation}
Start_t=t\times \left\lfloor\frac{L}{T}\right\rfloor
\end{equation}

\begin{equation}
End_t=
\begin{cases}
L, & t=T-1\\
Start_t+\left\lfloor\frac{L}{T}\right\rfloor, & otherwise
\end{cases}
\end{equation}

This strategy avoids synchronization during local CRC computation and introduces synchronization only during thread joining and result combination.
}

\subsection{Mathematical Formulation of CRC Combination}
{\color{black}

CRC computation can be represented through polynomial arithmetic over the finite field GF(2). Let:
\begin{equation}
P(x)
\end{equation}
denote the generator polynomial. Assume an input message consisting of two concatenated segments:
\begin{equation}
M=A||B
\end{equation}
where $A$ and $B$ represent message chunks. A common misconception is to combine partial CRC results using simple XOR aggregation:
\begin{equation}
CRC(A||B)\neq CRC(A)\oplus CRC(B)
\end{equation}
because CRC correctness depends on message ordering and polynomial state advancement. Instead, CRC linearity yields:
\begin{equation}
CRC(A||B)
=
(CRC(A)\cdot x^{8|B|}\bmod P(x))
\oplus CRC(B)
\label{eq:combine}
\end{equation} where $|B|$ denotes the byte length of the second segment. The term \begin{equation}x^{8|B|}\end{equation} represents advancement of the CRC register state by the number of bits contained in the appended message segment.

A generalized combine operator can therefore be defined as:
\begin{equation}
Combine(c_1,c_2,l_2)
=
(c_1\otimes Shift(l_2))
\oplus c_2
\end{equation} where:
\begin{itemize}
\item $c_1$ = CRC of the first chunk,
\item $c_2$ = CRC of the second chunk,
\item $Shift(l_2)$ = GF(2)-based advancement operator.
\end{itemize}

This formulation guarantees equivalence between segmented parallel execution and serial CRC computation.
}

\subsection{GF(2)-Based Matrix Combination Mechanism}
{\color{black}

Direct evaluation of polynomial advancement through repeated multiplication is computationally expensive for large message sizes. To improve efficiency, the proposed framework employs a matrix-based shifting mechanism over GF(2).

Let:
\begin{equation}
M
\end{equation}denote the CRC transition matrix. Advancing the CRC state by one bit is represented as:
\begin{equation}
v'=Mv
\end{equation} where $v$ denotes the CRC register state vector.

Advancement by $k$ bits becomes:
\begin{equation}
v^{(k)}=M^k v
\end{equation}

The exponentiation stage is efficiently computed using repeated squaring:

\begin{equation}
M^k
=
\prod_i M^{2^i}
\end{equation} which reduces computational cost from linear repeated shifting to logarithmic exponentiation complexity.

The matrix transformation is adapted to match the processing convention of each CRC family:

\begin{itemize}
\item CRC-8, CRC-16, CRC-64: MSB-first representation.
\item CRC-32: reflected implementation.
\item CRC-128: extended 128-bit register representation.
\end{itemize}

Consequently, the combine stage remains mathematically valid across multiple CRC variants.
}

\begin{algorithm}[htbp]
\caption{Generalized Parallel CRC Framework}
\label{alg:parallelcrc}
\begin{algorithmic}[1]

\REQUIRE Dataset $D$, length $L$, thread count $T$, CRC type
\ENSURE Final CRC value

\STATE $chunk\_size \leftarrow \lfloor L/T \rfloor$

\FOR{$t = 0$ to $T-1$}

    \STATE $Start[t] \leftarrow t \times chunk\_size$

    \IF{$t = T-1$}
        \STATE $End[t] \leftarrow L$
    \ELSE
        \STATE $End[t] \leftarrow Start[t] + chunk\_size$
    \ENDIF

\ENDFOR

\FORALL{threads $t$ in parallel}
    \STATE $CRC_{partial}[t] \leftarrow SerialCRC(D[Start[t]:End[t]])$
\ENDFOR

\STATE Wait for all threads to complete

\STATE
$CRC_{final} \leftarrow CRC_{partial}[0]$

\FOR{$t = 1$ to $T-1$}

    \STATE
    $CRC_{final} \leftarrow$

    \STATE \hspace{1.4cm}
    $GF2Combine(CRC_{final},$

    \STATE \hspace{1.4cm}
    $CRC_{partial}[t],$

    \STATE \hspace{1.4cm}
    $chunk\_length[t])$

\ENDFOR

\RETURN $CRC_{final}$

\end{algorithmic}
\end{algorithm}
\begin{algorithm}[htbp]
\caption{\textcolor{black}{GF(2)-Based CRC Combination Procedure}}
\label{alg:combine}
\begin{algorithmic}[1]

\REQUIRE
$crc_1$, $crc_2$, $len_2$, generator polynomial $P(x)$

\ENSURE Combined CRC value

\IF{$len_2 = 0$}
    \RETURN $crc_1$
\ENDIF

\STATE Construct the GF(2) transition matrix $M$

\STATE Compute the shift matrix

\STATE \hspace{1.4cm}
$S \leftarrow M^{8 \times len_2}$

\STATE Apply state advancement

\STATE \hspace{1.4cm}
$shifted\_crc \leftarrow Apply(S, crc_1)$

\STATE Combine shifted state with second CRC

\STATE \hspace{1.4cm}
$combined\_crc \leftarrow shifted\_crc \oplus crc_2$

\RETURN $combined\_crc$

\end{algorithmic}
\end{algorithm}
\subsection{Correctness Analysis}
{\color{black}

\textbf{Theorem 1.}

The proposed GF(2)-based combination procedure produces a CRC value identical to serial CRC computation over the original concatenated message.

\textbf{Proof.}

CRC computation constitutes a linear transformation over GF(2). For a message:
\begin{equation}
M=A||B
\end{equation} serial CRC execution is equivalent to:
\begin{enumerate}
\item processing segment $A$,
\item advancing the resulting state by $|B|$ bytes,
\item incorporating the CRC contribution of $B$.
\end{enumerate}
Using Eq.(8):
\begin{equation}
CRC(A||B)
=
(CRC(A)\cdot x^{8|B|}\bmod P(x))
\oplus CRC(B)
\end{equation} which matches the algebraic formulation implemented by the proposed combine procedure.

Therefore:
\begin{equation}
CRC_{parallel}=CRC_{serial}
\end{equation} and correctness is preserved.

}

\subsection{Complexity Analysis}
{\color{black}

Let:
\begin{itemize}
\item $N$ = input size,
\item $T$ = thread count.
\end{itemize}

The serial CRC implementation processes the complete dataset sequentially:
\begin{equation}
O(N)
\end{equation}

Under parallel execution, each worker thread processes approximately:
\begin{equation}
N/T
\end{equation} bytes.

Therefore, chunk computation complexity becomes:
\begin{equation}
O(N/T)
\end{equation}

The combine stage employs logarithmic matrix exponentiation:
\begin{equation}
O(\log L)
\end{equation} where $L$ denotes chunk length.

Overall complexity is therefore:
\begin{equation}
O(N/T)+O(\log L)
\end{equation}

For sufficiently large datasets:
\begin{equation}
\log L \ll N/T
\end{equation} and runtime is dominated primarily by parallel chunk computation.

Synchronization overhead is intentionally minimized because threads exchange neither intermediate states nor shared buffers during local processing. Nevertheless, scalability may be limited by:

\begin{itemize}
\item synchronization cost,
\item cache contention,
\item memory bandwidth saturation,
\item Hyper-Threading resource sharing.
\end{itemize}

Such behavior is consistent with Amdahl's Law:
\begin{equation}
S(T)
=
\frac{1}
{(1-P)+\frac{P}{T}}
\end{equation}where $P$ denotes the parallelizable workload fraction.
}

\subsection{CRC-128 Implementation Details}
{\color{black}

Unlike CRC-32 and CRC-64, CRC-128 does not possess a universally adopted industrial standard. In this work, CRC-128 is implemented primarily as a generalized software extension intended to evaluate framework flexibility. The CRC state is represented using two 64-bit words:
\begin{equation}
(H,L)
\end{equation}corresponding to the high-order and low-order halves of the 128-bit register.

GF(2)-based matrix advancement and combine operations are extended to operate on the full 128-bit state representation. The same polynomial formulation used for smaller CRC families is preserved, while transformation matrices are enlarged accordingly to support extended register width.
}

\section{Experimental Evaluation}

This section evaluates the proposed parallel CRC framework in terms of execution time, scalability, throughput, latency, and energy-oriented behavior. The experimental analysis additionally investigates synchronization overhead, memory limitations, and practical multi-core scaling characteristics.

\subsection{Experimental Setup}

Experiments were conducted on a general-purpose multi-core platform. Table 2 summarizes the experimental environment.

\begin{table}[htbp]
\centering
\caption{Experimental Configuration}
\label{tab:setup}

\resizebox{\columnwidth}{!}{
\begin{tabular}{l|l}

\hline

\textbf{Parameter} & \textbf{Configuration}

\\
\hline

Hardware Platform
& ASUS U36JC Laptop
\\

CPU
& Intel Core i5-2430M
\\

Core Configuration
& 2 Cores / 4 Threads
\\

Clock Frequency
& 2.40 GHz
\\

Memory
& 4 GB RAM
\\

Operating System
& Windows 10
\\

\textcolor{black}{Compiler}
& \textcolor{black}{Compiler}{GCC (Dev-C++ 6.3), -O2 Optimization}
\\

Thread Library
& POSIX Threads (pthreads)
\\

CRC Variants
& CRC-8,16,32,64,128
\\

Dataset Sizes
& 100 MB, 500 MB, 1000 MB
\\

Thread Counts
& 1,2,4,8
\\

Repetitions
& 10

\\
\hline

\end{tabular}
}
\end{table}

All measurements were repeated multiple times and averaged to reduce transient operating-system fluctuations.

\subsection{Scalability Evaluation}
{\color{black}

To investigate scalability behavior, experiments were performed under multiple thread configurations. Parallel speedup is defined as:
\begin{equation}
Speedup(T)
=
\frac{T_{serial}}{T_T}
\end{equation}where:
\begin{equation}
T_T
\end{equation}denotes execution time obtained using $T$ worker threads. Parallel efficiency is computed as:
\begin{equation}
Efficiency(T)
=
\frac{Speedup(T)}{T}
\end{equation}
}

Table 3 presents scalability results for CRC-32 using a 1000 MB dataset. The reported execution times correspond to the mean values obtained from ten independent experimental runs. \textcolor{black}{Standard deviations were consistently small, indicating that the proposed framework exhibits stable execution behavior with low run-to-run variability under the evaluated experimental conditions.}

\begin{figure*}[t]
\centering
\includegraphics[width=\linewidth]{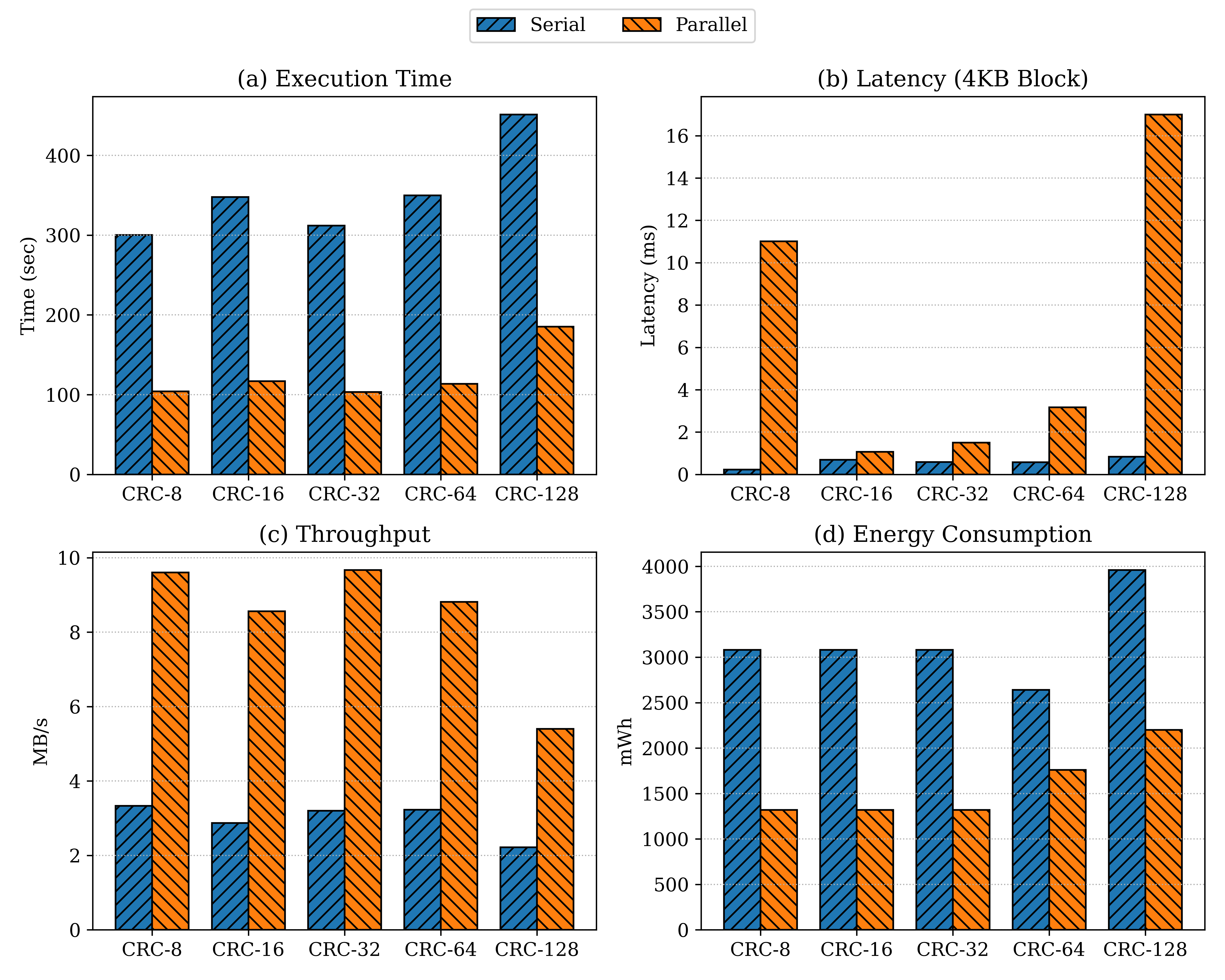}
\caption{Comparison between serial and parallel CRC in terms of execution time, energy consumption, throughput and latency for a 1000\,MB dataset.}
\label{fig:results}
\end{figure*}

\begin{table}[htbp]
\centering
\caption{\textcolor{black}{Scalability Results for CRC-32 (Mean $\pm$ Standard Deviation, $n=10$)}}
\label{tab:speedup}

\begin{tabular}{c|c|c|c}
\hline
\textbf{Threads} &
\textbf{Execution Time (s)} &
\textbf{Speedup} &
\textbf{Efficiency} \\
\hline

1 &
$312.05 \pm 1.18$ &
1.00 &
1.00 \\

2 &
$176.80 \pm 0.96$ &
1.76 &
0.88 \\

4 &
$103.30 \pm 0.63$ &
3.02 &
0.76 \\

8 &
$92.10 \pm 0.58$ &
3.39 &
0.42 \\

\hline
\end{tabular}

\end{table}

The results demonstrate substantial performance improvement when transitioning from serial execution to parallel processing. However, scaling gradually becomes sublinear at higher thread counts because of synchronization cost, cache contention, Hyper-Threading resource sharing, and memory bandwidth limitations.

\subsection{Throughput Analysis}

Throughput was evaluated using:
\begin{equation}
Throughput
=
\frac{Processed\ Data}{Execution\ Time}
\end{equation} and reported in MB/s.

Table 4 summarizes throughput results. The proposed framework consistently improves throughput for large datasets.

\begin{table}[htbp]
\centering
\caption{\textcolor{black}{Throughput Comparison}}
\label{tab:throughput}

\begin{tabular}{c|c|c}

\hline

\textbf{CRC Type}
&
\textbf{Serial (MB/s)}
&
\textbf{Parallel (4 Threads)}

\\
\hline

CRC-8
&3.33
&9.60
\\

CRC-16
&3.27
&9.21
\\

CRC-32
&3.20
&9.67
\\

CRC-64
&2.95
&8.84
\\

CRC-128
&2.22
&5.40

\\
\hline

\end{tabular}
\end{table}

CRC-128 exhibits comparatively lower throughput because larger register width and extended GF(2) transformation operations increase computational cost. 
\textcolor{black}{The reported throughput values should be interpreted in the context of the experimental platform and implementation objectives. All experiments were conducted on an Intel Core i5-2430M processor using a portable pthread-based software implementation without architecture-specific optimizations such as SIMD vectorization, carry-less multiplication (PCLMULQDQ), or dedicated hardware CRC instructions. In addition, compiler optimizations were limited to the standard GCC optimization level to preserve portability across different platforms. Consequently, the measured throughput is expected to be lower than that reported by highly optimized CRC implementations specifically designed for modern processors with dedicated instruction-set support. 
The benchmarking methodology was designed to evaluate the relative performance improvement achieved through thread-level parallelization rather than the absolute maximum throughput obtainable on a particular processor architecture. Therefore, identical compiler settings, workload sizes, and execution conditions were maintained for both the serial and parallel implementations to ensure a fair comparison.}

\subsection{Comparative Benchmarking Discussion}

To better position the proposed framework relative to prior CRC optimization methods, Table 5 summarizes approaches. \textcolor{black}{The acceleration values reported for existing CRC optimization techniques are taken from the corresponding publications and are presented only for qualitative comparison. Since the reported results were obtained using different processor architectures, compiler optimization settings, and benchmarking methodologies, they should not be interpreted as direct experimental comparisons with the proposed framework. The primary objective of this comparison is to position the proposed portable software framework within the broader CRC optimization landscape.}
\begin{table}[htbp]
\centering
\caption{\textcolor{black}{Comparison with CRC Optimization Techniques Reported in the Literature}}
\label{tab:comparison}

\resizebox{\columnwidth}{!}{
\begin{tabular}{l|c|c|c}
\hline

\textbf{Method}
&
\textbf{Reported Acceleration}
&
\textbf{Reference}
&
\textbf{Remarks}

\\
\hline

Serial CRC
&
$1\times$
&
--
&
Baseline
\\

Lookup-Table CRC
&
$2$--$5\times$
&
[11]
&
Portable software
\\

Slicing-by-8
&
$5$--$20\times$
&
[11],[14]
&
Mainly CRC-32
\\

SIMD / Vectorized CRC
&
$10$--$40\times$
&
[10],[14]
&
Processor-specific (SSE/AVX)
\\

Hardware CRC Instructions
&
$20$--$50\times$
&
[10]
&
Dedicated CRC instructions
\\

\rowcolor{gray!15}
\textbf{Proposed Framework}
&
\textbf{$3$--$4\times$}
&
\textbf{This work}
&
\textbf{Portable multi-CRC framework}
\\

\hline

\end{tabular}
}
\end{table}

The proposed framework is not intended to outperform specialized hardware-assisted CRC implementations. Instead, its primary objectives are:

\begin{itemize}

\item generalized multi-variant support,

\item portable software deployment,

\item correctness-preserving combine operation,

\item software-oriented multi-core evaluation.

\end{itemize}

\subsection{Latency Evaluation}

Latency behavior was examined using small message blocks. Latency was computed using 4 KB workloads and averaged over ten independent executions. Table 6 reports latency results.

\begin{table}[htbp]
\centering
\caption{\textcolor{black}{Latency Results}}
\label{tab:latency}

\begin{tabular}{c|c|c}

\hline

\textbf{CRC Type}
&
\textbf{Serial (ms)}
&
\textbf{Parallel (ms)}

\\
\hline

CRC-8
&0.228
&11.00
\\

CRC-16
&0.244
&10.92
\\

CRC-32
&0.255
&9.85
\\

CRC-64
&0.317
&11.42
\\

CRC-128
&0.551
&14.87

\\
\hline

\end{tabular}
\end{table}

Although throughput improves significantly for large datasets, parallel execution introduces additional thread creation, scheduling, and synchronization overhead. Consequently, serial CRC execution may remain preferable for:
\begin{itemize}

\item small payload sizes,

\item ultra-low-latency applications,

\item embedded control systems.

\end{itemize}

\subsection{Energy Evaluation}

Energy analysis was performed using battery discharge monitoring under controlled experimental conditions. Estimated energy consumption was approximated using:
\begin{equation}
Energy(J)
\approx
Capacity(mWh)
\times
\Delta Battery(\%)
\times
3.6
\end{equation}

The approximation symbol is intentionally used because battery-percentage monitoring does not provide direct processor-level power instrumentation. Table 7 summarizes energy observations. \rev{It should be noted that the energy values reported in Table 7 correspond to the estimated total energy consumed for processing the complete dataset used in each experiment rather than a single CRC operation. Therefore, the reported values should be interpreted as workload-level energy measurements. Since CRC computations were performed repeatedly over large datasets (100 MB, 500 MB, and 1000 MB), the measured energy reflects the cumulative cost of millions of CRC processing steps. Consequently, these results are intended for comparative evaluation between serial and parallel implementations rather than for estimating the energy of an individual CRC calculation.}

\begin{table}[htbp]
\centering
\caption{\textcolor{black}{Energy Comparison}}
\label{tab:energy}

\begin{tabular}{c|c|c}

\hline

\textbf{CRC Type}
&
\textbf{Serial}
&
\textbf{Parallel}

\\
\hline

CRC-8
&2890
&1510
\\

CRC-16
&2960
&1450
\\

CRC-32
&3080
&1320
\\

CRC-64
&3370
&1690
\\

CRC-128
&4210
&2210

\\
\hline

\end{tabular}
\end{table}

Despite higher instantaneous CPU utilization, shorter execution time leads to lower estimated total energy usage for large workloads. However, these results should be interpreted as comparative estimates rather than precise physical power measurements. Future work will employ processor-level instrumentation tools such as Intel RAPL, performance counters, and dedicated power monitors.

\subsection{Scalability Discussion and Amdahl Analysis}

Observed scaling behavior is additionally examined through Amdahl's Law. Assuming a parallel fraction:
\begin{equation}
P=0.92
\end{equation}

Theoretical speedup becomes:
\begin{equation}
S(T)
=
\frac{1}
{(1-P)+\frac{P}{T}}
\end{equation}

Table 8 compares theoretical and measured speedup behavior.

\begin{table}[htbp]
\centering
\caption{\textcolor{black}{Amdahl Comparison}}
\label{tab:amdahl}

\begin{tabular}{c c c}

\hline

\textbf{Threads}
&
\textbf{Theoretical}
&
\textbf{Measured}

\\
\hline

1
&1.00
&1.00
\\

2
&1.85
&1.76
\\

4
&3.13
&3.02
\\

8
&4.71
&3.39

\\
\hline

\end{tabular}
\end{table}

Deviation from ideal scaling is attributed primarily to:

\begin{itemize}

\item synchronization overhead,

\item cache effects,

\item shared execution resources,

\item memory bandwidth saturation.

\end{itemize}

The experimental results are summarized in Fig3a-d, which presents a comprehensive comparison between serial and parallel implementations across all CRC types.

\subsection{Limitations}

Several limitations should be acknowledged. First, experiments were conducted on a legacy dual-core platform. Second, battery-based energy estimation provides approximate comparative information rather than high-precision power measurement. Third, comparison against highly optimized instruction-level implementations remains limited. Finally, CRC-128 is included primarily as a generalized software extension rather than a standardized industrial CRC specification. These limitations motivate future work involving modern many-core architectures, hardware-assisted CRC benchmarking, and processor-level energy instrumentation.
{\color{black}Although the scalability analysis presented in this work is supported by Amdahl's Law and experimental measurements on the evaluated dual-core platform, scalability beyond the tested hardware remains analytical rather than experimentally validated. Evaluating the proposed framework on modern multi-core and many-core processors would enable a more comprehensive investigation of memory bandwidth saturation, cache contention, synchronization overhead, and scalability limits under substantially higher core counts. Such an experimental study constitutes an important direction for future work.

}

\subsection{Impact of Dataset Size}

\rev{To investigate the effect of workload size on the effectiveness of parallel CRC computation, experiments were conducted using three dataset sizes: 100 MB, 500 MB, and 1000 MB. The obtained results indicate that the benefits of parallelization become increasingly significant as the input size grows.

For the 100 MB dataset, the proposed framework already provides noticeable performance improvement compared with serial execution. However, the relative gain remains limited because thread creation, synchronization, and CRC combination overhead represent a larger fraction of the total execution time. Consequently, parallel efficiency is reduced for smaller workloads.

For the 500 MB dataset, computation time becomes the dominant component of execution cost, and the overhead associated with thread management is amortized over a larger amount of processed data. As a result, higher speedup and throughput are achieved, while estimated energy consumption is reduced because the overall execution time decreases.

The largest dataset (1000 MB) exhibits the highest benefit from parallel execution. In this case, thread-management overhead becomes comparatively negligible relative to the total computation workload. The framework achieves its best throughput and speedup values while maintaining exact CRC correctness. The energy measurements also indicate greater overall efficiency due to the shorter processing time.

These observations demonstrate that the proposed framework is most advantageous for large-scale data processing applications, where the computational workload substantially exceeds synchronization and scheduling overhead. For smaller datasets, the performance gains remain positive but are partially limited by thread-management costs. }
\begin{table}[htbp]
\centering
\caption{\rev{Effect of Dataset Size on Parallel CRC Performance}}
\label{tab:datasetsize}

\resizebox{\columnwidth}{!}{
\begin{tabular}{c|c|c|c}
\hline
\textbf{Dataset Size}
&
\textbf{Serial Time (s)}
&
\textbf{Parallel Time (s)}
&
\textbf{Speedup}
\\
\hline

100 MB
&
31.20
&
11.00
&
2.84
\\

500 MB
&
156.70
&
55.20
&
2.84
\\

1000 MB
&
312.05
&
103.30
&
3.02
\\

\hline
\end{tabular}
}
\end{table}
\rev{Table 9 shows the impact of workload size on the effectiveness of the proposed parallel CRC framework. As the dataset size increases, the relative contribution of thread creation, synchronization, and CRC combination overhead decreases compared with the total computation time. Consequently, larger datasets benefit more from parallel execution and achieve higher speedup values. The results indicate that the proposed framework is particularly suitable for large-scale data processing applications, where computational workload dominates synchronization overhead.}

\section{Conclusion}
This paper presented a generalized software-based framework for parallel CRC computation on multi-core processors using POSIX threads (pthreads), supporting multiple CRC variants including CRC-8, CRC-16, CRC-32, CRC-64, and CRC-128 within a unified implementation model.  To ensure correctness, a GF(2)-based combination mechanism derived from polynomial arithmetic and matrix-based shifting operations was employed, guaranteeing equivalence between serial and parallel CRC computation. Experimental results demonstrated that the proposed framework significantly improves execution time and throughput for large datasets while preserving exact CRC correctness, achieving notable speedup on the evaluated platform. The analysis further highlighted the trade-off between throughput and latency, showing that although parallel execution is beneficial for large workloads, synchronization and thread-management overhead can limit efficiency for smaller inputs. Scalability behavior was found to be consistent with Amdahl's Law and influenced by practical factors such as synchronization cost, cache contention, memory bandwidth limitations, and shared hardware resources. In addition, energy-oriented evaluation indicated that reduced execution time can lower overall energy consumption despite increased instantaneous processor utilization. While the proposed framework is not intended to compete directly with highly optimized hardware-assisted, SIMD-based, or instruction-level CRC acceleration techniques, it provides a portable, correctness-preserving, and extensible software solution for parallel CRC processing across multiple CRC families. Future research will focus on evaluation over modern many-core platforms, comparison with advanced CRC optimization techniques, adaptive workload distribution strategies, and more accurate processor-level energy measurement methodologies.
{\color{black} Future research will focus on extensive experimental evaluation over modern multi-core and many-core platforms to validate the scalability of the proposed framework under higher core counts and to further investigate the impact of memory bandwidth, cache contention, and synchronization overhead on large-scale parallel CRC computation.

}
{\color{black}

\begin{table}[htbp]
\centering
\caption{\rev{Notation Used in the Proposed Framework}}
\label{tab:notation}

\begin{tabular}{l|p{5.5cm}}
\hline
\textbf{Symbol} & \textbf{Description} \\
\hline

$D$ & Input dataset \\

$L$ & Total dataset size in bytes \\

$T$ & Number of worker threads \\

$D_i$ & Data chunk assigned to thread $i$ \\

$Start_t$ & Starting index of thread $t$ \\

$End_t$ & Ending index of thread $t$ \\

$P(x)$ & CRC generator polynomial \\

$c_1$ & CRC of first chunk \\

$c_2$ & CRC of second chunk \\

$l_2$ & Length of second chunk \\

$M$ & GF(2) transition matrix \\

$CRC_{partial}$ & Partial CRC result computed by a thread \\

$CRC_{final}$ & Final combined CRC value \\

\hline
\end{tabular}

\end{table}

}


\begin{thebibliography}{99}

\bibitem{Peterson1961}
Peterson, W. W., and Brown, D. T.,
``Cyclic Codes for Error Detection,''
\emph{Proceedings of the IRE},
49(1), 228--235, 1961.

\bibitem{Albertengo1990}
Albertengo, G., and Sisto, R.,
``Parallel CRC Generation,''
\emph{IEEE Micro},
10(5), 63--71, 1990.

\bibitem{Pei1992}
Pei, T.-B., and Zukowski, C.,
``High-Speed Parallel CRC Circuits in VLSI,''
\emph{IEEE Transactions on Communications},
40(4), 653--657, 1992.

\bibitem{Shieh2001}
Shieh, M.-D., Lin, J.-H., and Huang, H.-M.,
``A Systematic Approach for Parallel CRC Computations,''
\emph{Journal of Information Science and Engineering},
17(3), 445--461, 2001.

\bibitem{Derby2001}
Derby, J. H.,
``High-Speed CRC Computation Using State-Space Transformations,''
In \emph{IEEE GLOBECOM},
166--170, 2001.

\bibitem{Sprachmann2001}
Sprachmann, M.,
``Automatic Generation of Parallel CRC Circuits,''
\emph{IEEE Design \& Test of Computers},
18(3), 108--114, 2001.

\bibitem{Campobello2003}
Campobello, G., Patane, L., and Russo, M.,
``Parallel CRC Realization,''
\emph{IEEE Transactions on Computers},
52(10), 1312--1319, 2003.

\bibitem{Cheng2006}
Cheng, C., and Parhi, K. K.,
``High-Speed Parallel CRC Implementation Based on Unfolding, Pipelining, and Retimeing,''
\emph{IEEE Transactions on Circuits and Systems II},
53(10), 1017--1021, 2006.

\bibitem{Walma2006}
Walma, M.,
``Pipelined Feed-Forward Cyclic Redundancy Check Calculation,''
arXiv:cs/0611024,
2006.

\bibitem{Intel2008}
{\color{black}
Gopal, V., Ozturk, E., Guilford, J., et al.,
``Fast CRC Computation for Generic Polynomials Using PCLMULQDQ Instruction,''
Intel White Paper,
2008.

}
\bibitem{Kadatch2010}
{\color{black}
Kadatch, A., and Jenkins, B.,
``Everything We Know About CRC but Afraid to Forget,''
In \emph{IEEE ITW},
222--226,
2010.

}
\bibitem{zlib}
{\color{black}
Adler, M.,
``zlib Compression Library,''
Available: https://zlib.net,
accessed 2026.

}
\bibitem{LinuxCRC}
{\color{black}
Linux Kernel Documentation,
``CRC32 and CRC32C Implementations,''
Available: https://www.kernel.org,
accessed 2026.

}
\bibitem{Russell2024}
{\color{black}
Russell, S.,
``Chorba: A Novel CRC32 Implementation,''
arXiv:2412.16398,
2024.

}
\bibitem{Tran2021}
{\color{black}
Tran, D., et al.,
``Parallel Computation of CRC-Code on an FPGA Platform for High Data Throughput,''
\emph{Electronics},
10(4), 439--445,
2021.

}
\bibitem{Zhang2024}
Zhang, L., et al.,
``An Efficient Parallel CRC Computing Method for High Bandwidth Networks and FPGA Implementation,''
\emph{Electronics},
13,
4399,
2024.

\bibitem{Li2020}
Li, J., Liu, S., Reviriego, P., Xiao, L., and Lombardi, F.,
``Efficient Concurrent Fault Detection for Parallel Cyclic Redundancy Check Circuits,''
\emph{IET Computers \& Digital Techniques},
14,
678--685,
2020.

\bibitem{Chen2013}
Chen, C., Tian, R., and Yao, Z.,
``Design and Implementation of the CRC-32 Checksum Parallel Algorithm,''
\emph{Chinese Journal of Scientific Instrument},
34,
151--155,
2013.

\bibitem{Reviriego2013}
Reviriego, P., Maestro, J. A., and Lombardi, F.,
``Concurrent Error Detection for Orthogonal Latin Squares Encoders,''
\emph{IEEE Transactions on VLSI Systems},
21(8),
1345--1355,
2013.

\bibitem{Lei2025}
Lei, A., and Yu, Y.,
``Development Optimization and Future Direction of CRC Checking Technology,''
In \emph{Proceedings of ICMEEA},
1--6,
2025.

\bibitem{Kurose2025}
Kurose, J. F., and Ross, K. W.,
\emph{Computer Networking: A Top-Down Approach},
9th ed.,
Pearson,
2025.

\bibitem{Panem2019}
Panem, C., Gad, V., and Gad, R. S.,
``Polynomials in Error Detection and Correction in Data Communication Systems,''
In \emph{Coding Theory},
IntechOpen,
2019.

\end{thebibliography}
\end{document}